\documentclass[%
 reprint,
 amsmath,amssymb,
 aps,
]{revtex4-2}

\usepackage[export]{adjustbox}
\usepackage{relsize}
\usepackage{graphicx}
\usepackage{bm}
\usepackage{graphicx}
\usepackage{comment}
\usepackage{float}
\usepackage[caption=false]{subfig}
\usepackage{rotating}
\usepackage[]{hyperref}
\hypersetup{
colorlinks,
linkcolor={blue},
linktocpage=true,
citecolor={red},
urlcolor={green}
}

\newcommand{\MuToEmConv}{$\mu^-N \rightarrow e^- N$}
\begin{document}

\title{Experimental Measurements of the Muon g - 2 and Searches for Charged Lepton Flavor Violation in the Muon Sector}
\author{Sophie Middleton}
\date{\today}

\begin{abstract}
\textbf{Contribution to the 25th International Workshop on Neutrinos from Accelerators}
Since its discovery, the muon has proven to be an invaluable probe of the Standard Model (SM). Muons are readily available in tertiary beams in facilities around the world. They do not decay hadronically and have a lifetime of a few $\mu$ s; consequently, muon experiments offer clean, high-statistics environments to make precision measurements and search for new physics that could appear through deviations from the SM expectation. The 2020s have seen a renaissance in muon physics highlighted by the high-profile results from the Fermilab Muon $g - 2$ experiment which continues to provide successive measurements of the muon's anomalous magnetic moment with world-leading precision. In addition, a suite of experiments is coming online to search for new physics in the form of charged lepton flavor violation in the muon sector. These experiments will probe effective mass scales of new physics up to $10^4$ TeV/c$^2$, far beyond the reach of direct searches at colliders. This article explores the motivations, recent results, and status of these experiments.
\end{abstract}

\maketitle

\section{The Muon}

The muon was first observed in a laboratory in 1936 \cite{anderson} and was eventually proven to be a unique particle, leading to the now infamous exclamation from Rabi: ``who ordered that?." This was a revolutionary discovery, the first evidence for generations of particles within the Standard Model (SM). Since this discovery, the muon has played a key role in helping us understand physics. Precision measurements of muon decay provide unparalleled insight into the SM. Studies of muon decay both determine the overall strength and establish the chiral structure of weak interactions. The muon's fundamental properties provide experimental advantages. Its mass is $\sim$ 105 MeV/c$^{2}$ meaning that the muon does not decay hadronically; consequently, muon experiments provide very clean environments, with well-understood SM backgrounds. Muons can be created in large numbers as tertiary beams emanating from pion decay as a result of protons on fixed targets. Furthermore, the muon has a lifetime, at rest, of $\sim$ 2.2 $\mu$s, long enough that they can be transported and stored, accumulating high statistics in beams and projected on to targets,  but short enough that the decay itself can be utilized. High-precision measurements of the muon's anomalous magnetic moment, $a_{\mu}$, offer sensitivity to the completeness of the SM and beyond SM (BSM) theories that predict deviations from SM expectation. Muon charged lepton flavor violation (CLFV) channels probe a range of well-motivated new physics scenarios. Due to the high-intensity muon beams available at facilities such as PSI, Fermilab and J-PARC, current and projected constraints on muon CLFV channels are orders of magnitude beyond analogous constraints on tau channels. The next generation of muon CLFV experiments will probe effective masses to $10^{4}$ TeV/c$^2$ \cite{bernstein_history}. The present decade is an unprecedented era for muon physics, with experiments at PSI, Fermilab, J-PARC and CERN all planning to take data to either enhance understanding of the muon's anomalous magnetic moment or search for new physics using high-intensity muon beams. This article explores the status of these current and upcoming experiments.

\section{Measuring the Muon $g -2$}

The anomalous magnetic moment of the muon, $a_{\mu}$, is defined as $a_{\mu} = \frac{g_{\mu} - 2}{2}$. Precision measurements of $a_{\mu}$ are a stringent probe of the SM and any deviations of this value from theoretical predictions could be a sign of new physics.


The Fermilab Muon $g - 2 $ Experiment follows an experimental approach used by its predecessor at BNL \cite{BNL}, and originally pioneered at CERN \cite{CERN-Mainz-Daresbury:1978ccd}. A 3.1 GeV/c, spin-polarized, $\mu^{+}$ beam orbits within a 7.1 m radius storage ring with a 1.45T uniform magnetic field. The muon spin precesses in the highly uniform field ($\vec{B}$). The relative precession frequency of the spin with respect to the momentum, denoted as the anomalous precession frequency,  $\omega_{a}$, is given as the difference between the spin precession frequency, $\omega_{s}$, and the cyclotron frequency, $\omega_c$:

\begin{equation}\omega_a = \omega_s - \omega_c = a_{\mu}  \frac{q \vec{B}}{m},
\end{equation} in the the absence of an electric field and in the limit of planar muon orbits. A  measurement of the anomalous precession frequency, coupled with precise knowledge of the storage ring magnetic field, provides a direct probe of the anomalous magnetic moment.

\begin{figure}[ht!]
\centering{
\includegraphics[width=0.5\textwidth]{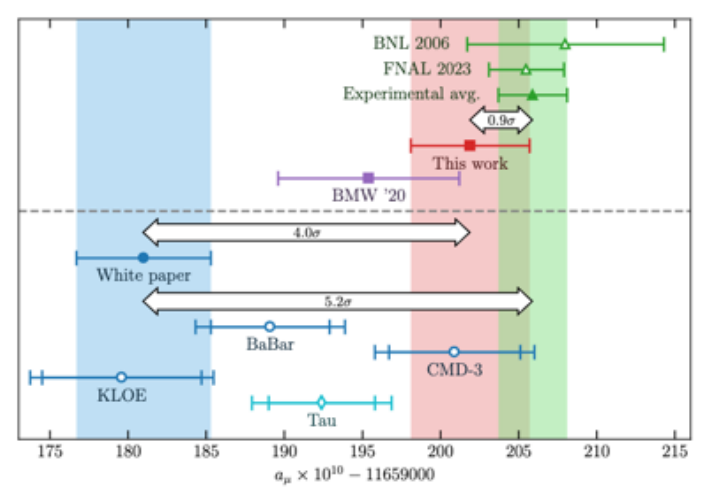}
\caption{Comparison between experimental measurements of muon g --2 and theoretical predictions from \cite{Boccaletti:2024guq}. \label{fig:g2results}}}
\end{figure}

Muons decay to produce positrons and neutrinos. The decay positron flux becomes a means to understand the precession frequency. Parity violation in the weak force means that the highest energy positrons are emitted primarily along the direction of the muon's spin, in the rest frame.  There is a resulting modulation in the energy spectrum of the positrons in the lab frame. With precise measurement of the magnetic field and through measuring the positron flux around the ring, $a_{\mu}$ is derived. 

The Fermilab Muon $g - 2$ experiment took data from 2018 until 2023. During this time, data totalling more than $\times 21$ that at BNL experiment was acquired. So far, two results have been published, analyzing data from the first three runs, both present the most precise determination of $a_{\mu}$ at the date of publication: \cite{Muong-2:2021ojo,Muong-2:2023cdq}. The final three runs, making up over half the total data, are currently being analyzed. The combined result for the first three runs is $a_{\mu}$ =  1.16592055(24) $\times  10^{-11}$ (0.20 ppm) \cite{PhysRevD.110.032009}.  Figure \ref{fig:g2results} shows how this experimental measurement compares to various theory predictions for the SM contribution. There are large discrepancies of over 5 $\sigma$ compared to the SM prediction from the ``Muon g -- 2 Theory Initiative White Paper" of 2020 \cite{Aoyama_2020} which utilized a dispersive calculation to get the hadronic vacuum polarization contribution. However, a newer result presented in Ref. \cite{Boccaletti:2024guq}, which utilizes a hybrid approach that incorporates lattice calculations, sits much closer to the experimental result. Likewise, newer electronic scattering measurements, such as from CMD-3, have made theoretical predictions using dispersive approaches more challenging to interpret. It is evident that a large effort is needed to understand all these tensions and to provide insight into any potential deviation from SM expectation. Reference \cite{illinois} details various ongoing efforts from both theorists and experimentalists to resolve the tension in these calculations. 

The MUonE experiment \cite{Abbiendi:2016xup} aims to make a completely independent measurement of the hadronic contribution to $a_{\mu}$ using space-like scattering of 150 GeV muons from electrons in low-Z beryllium. MUonE will make precise measurements of the scattering angles as these muons traverse a series of 40 identical stations, each consisting of beryllium targets surrounded by silicon tracking layers. MUonE conducted a test beam in 2023 \cite{Driutti:2024vtg}. One tracking station was utilized, with a small-scale ECAL downstream and a target-less station upstream, to measure the incoming beam direction. Analysis of the data is still in progress. The full configuration with 40 stations could be prepared for the end of the decade, with a preceding stage with just 10 stations possibly taking place earlier. The final goal is to acquire $3.5 \times 10^{12}$ elastic scattering events with an electron energy larger than 1 GeV, over 3 years. This will allow a stat. error of 0.3$\%$ and 10 ppm sys. error. comparable with the results from the time-like dispersive approach and the lattice calculations.

In addition, the J-PARC g $-$ 2 experiment \cite{Iinuma:2011zz} aims to make a complementary measurement of $a_{\mu}$. The J-PARC experiment will employ a low-emittance, 300 MeV/c, muon beam, produced by re-acceleration of thermal muons regenerated by the laser resonant ionization of muonium atoms emitted from a silica aerogel \cite{Kamioka:2023xob}. This removes the need for strong focusing by an electric field which provides a significant correction term in the storage ring experiments. The muons are stored within a ring $20 \times$ smaller than the FNAL ring and silicon sensors are used to fully reconstruct the decay positrons. The J-PARC g -- 2 experiment plans to begin data taking in 2028. In 2024, the experiment showed the first evidence of muon cooling and subsequent re-acceleration. A systematic uncertainty on $\omega_a$ of 70 ppb is expected, but statistical uncertainty of 450 ppb will limit the overall precision, estimated for a two year running period.

\section{Searches for Charged Lepton Flavor Violation}

Reference \cite{bernstein_history} describes the long history of searches for muon charged lepton flavor violation (CLFV). Searches for CLFV in  $\mu \rightarrow e \gamma$ began as soon as the muon was discovered. The presence of neutrino oscillation implies the existence of CLFV at loop level, resulting from neutrino oscillations in a loop, however the calculated rates are $\mathcal{O}(10^{-54})$, far beyond the reach of any conceivable experiment \cite{Marciano2008}. Any experimental observation of a CLFV process would be a true departure from the SM, and would conclusively demonstrate the presence of new physics. Many well-motivated extensions of the SM predict much higher rates of the CLFV processes which fall within the reach of the upcoming generation of CLFV experiments: COMET \cite{ 10.1093/ptep/ptz125}, Mu2e \cite{mu2eTDR} ( searching for \MuToEmConv~); MEG-II \cite{MEGII2018} (searching for  $\mu^{+} \rightarrow e^{+}  \gamma$); and  Mu3e \cite{Mu3e2013} (searching for $\mu^{+} \rightarrow e^{+} e^{-} e^{+}$ ). Such new physics models include: SO(10) super-symmetry \cite{Calibbi2014,Calibbi2006}, models of scalar lepto-quarks \cite{Arnold, heeck} and models with additional Higgs doublets \cite{Abe2017}. Searches for \MuToEmConv~ can also help place limits on lepton flavor violating Higgs decays, Ref.~\cite{harnik} suggests that Mu2e and COMET have sensitivity up to $BR (h \rightarrow \mu e)$ of $10^{-10}$. CLFV measurements can provide invaluable insight into the mechanism behind neutrino masses.  Different mass generating Lagrangians predict different rates of CLFV, determining the rates of multiple CLFV processes allows us to discriminate among models~\cite{HAMBYE201413}.

New physics contributions to CLFV can be divided into two categories: dipole interactions, where a virtual loop process produces the conversion, and contact interactions, where a new virtual particle is exchanged at tree level and couples to the fermions. Searches for radiative $\mu^{+} \rightarrow e^{+} \gamma$ decays are only sensitive to electromagnetic dipole interactions. On the other hand, experiments searching for \MuToEmConv~ and $\mu^{+} \rightarrow e^{+} e^{-} e^{+}$ are sensitive to both photonic or contact type interactions. While searches for $\mu^{+} \rightarrow e^{+}  \gamma$ provide powerful constraints on the dipole operators, \MuToEmConv~ conversion is the most sensitive observable to explore operators involving quarks \cite{deGouvea2013, Davidson:2022nnl}. 
The relative contributions to the different CLFV channels reflects the nature of the underlying physics model responsible \cite{calibbi2017charged}. It is, therefore, crucial to explore all three muon CLFV channels to understand the potential signature of new physics. In the event that CLFV signals are observed, the conversion rate is dependent on the atomic nucleus \cite{kitano2002,Cirigliano2009,Leo}. Therefore, a way to elucidate a conversion signal would be to measure the conversion rate in a material complementary to aluminum (the target material proposed in Mu2e and COMET).

\subsection{Current Limits and Projections}

Table~\ref{table:limits} details the current limits on each of the three muon CLFV channels along with the projected limits of future searches for these three processes. Of course, all three also have discovery potential; upper limits are shown here purely for comparison with previous searches. All the experiments listed are either online already (MEG-II) or due to begin physics data collection in the next few years. In the following sections, each channel is discussed in detail in terms of the physics signature, backgrounds, and experimental approach.

\begin{table*}[t!]
\centering
\caption{Current limits and future projections for the three muon CLFV channels.}
\begin{tabular}{c |c |c} 
\hline
Channel & Current Limit (90 $\%$ C.L) & Future Projections (90 $\%$ C.L)\\ [0.5ex] 
\hline\hline
$\mu^{+} \rightarrow e^{+}  \gamma$ & $3.1 \times 10^{-13}$ (MEG+MEG-II) \cite{MEGII:2023ltw} & $6 \times 10^{-14}$ (MEG-II) \cite{MEGII:2023ltw}\\
\hline
$\mu^{+} \rightarrow e^{+} e^{+} e^{-}$ & $< 10^{-12}$  (SINDRUM) \cite{SINDRUM:1987nra} & $5\times 10^{-15} $($2 \times 10^{-16})$(Mu3e Phase I (II)) \cite{Mu3eLimit}\\
\hline
$\mu^{-} N \rightarrow e^{-} N$&$7\times 10^{-13}$ (SINDRUM-II) \cite{Bertl2006} & $8 \times 10^{-15}$ ($\mathcal{O}(10^{-17})$) (COMET, Phase-I (II))  \cite{COMET:2018auw}\\
&& $ 6 \times 10^{-16}$ ($8 \times 10^{-17}$ ) (Mu2e Run-I (I+II)) \cite{SU2020}(\cite{mu2eTDR})\\
\hline \hline
\end{tabular}
\label{table:limits}
\end{table*}

\subsection{$\mu^{-} \rightarrow e^{+} \gamma$}

In this scenario, the new physics signature is a two-body, back-to-back, $e^+$ and $\gamma$ pair, coincident in time, consistent with a $\mu^+$ decay at rest ($E_{e,\gamma} = 52.83 $ MeV) and emanating from the stopping target. The dominant background is an accidental one resulting from the time-coincidence of a $e^+$ and $\gamma$ emanating from two independent SM processes. Radiative muon decays also present a rare background, when the outgoing neutrinos carry negligible momentum, resulting in an $e^+ \gamma$ pair that are coincident in time with similar momenta.

\begin{figure}[ht!]
\centering{
\includegraphics[width=0.45\textwidth]{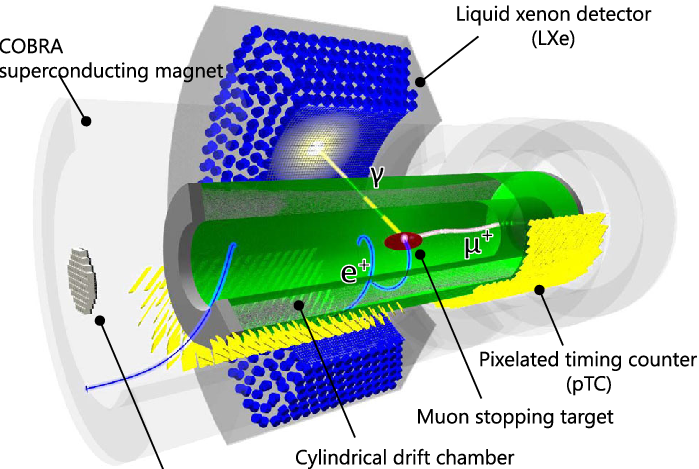}
\caption{A schematic of the MEG-II appratus. $\mu^{+}$ are incident on a thin target. The signal is a two body process involving a back-to-back photon and positron from the target \label{fig:MEG2}.}}
\end{figure}

The MEG-II Experiment \cite{MEGII:2021fah,MEGII2018} is based at PSI and is currently taking data. A schematic of the MEG-II apparatus is shown in Fig.~\ref{fig:MEG2}. The experiment utilizes an almost continuous $\mu^{+}$ beam with a rate of 3 $-$ 5 $\times 10^{7}$ $\mu$ per second. The muons are stopped on a thin target. The apparatus is optimized to exploit the two-body kinematics associated with the signal, looking for positrons and photons emanating from the target. A signal event would have a photon and positron that coincide in time, are back-to-back, and have equivalent energy of half the muon mass.

MEG-II employs a magnetic spectrometer to produce helical positron tracks. The COBRA (COnstant Bending RAdius) magnet has a maximum field of 1.3 T at the stopping target with a gradient that drops along the beam axis forcing signal-like positrons away from the target upstream or downstream along the beam axis in a helical trajectory. A cylindrical drift chamber (CDCH) is then used to reconstruct the positron tracks. The CDCH geometry allows for precise reconstruction of the positrons' position along the wire axis. The single-volume detector design is lighter than that of the MEG experiment, thus suppressing multiple scattering and improving momentum resolution. The positron time is measured in a pixilated timing counter array (pTC) with 512 pixilated timing counters. Each counter is made of scintillator with SiPM readout on the two ends. Each counter has a typical resolution of $\sim$ 100 ps resulting in excellent time resolution of $\sigma_t \sim $ 35 ps. The fitted track with information from the CDCH and SPX is propagated back to the stopping target for measurements of the positron position, direction, momentum, and time at the target. 

A single-volume, 800 L, liquid Xenon (LXe) calorimeter is used to detect the outgoing photons and measure their energy. The vertex of a photon/positron production is extrapolated back to the target.  A calorimeter containing 4096 multi-pixel photon counters (MPPCs) provides excellent position resolution. A radiative decay counter (RDC) is employed to eliminate some of the accidental time coincidences.

MEG-II started collecting physics data in 2021 following a successful engineering run. In 2023 their first physics result was released \cite{MEGII:2023ltw}. No signal was observed and a limit of $7.5 \times 10^{-13} $ at 90 $\%$ C.L was established from this data set. This was combined with the older MEG result to produce the limit shown in Table~\ref{table:limits} ($3.1 \times 10^{-13}$ at 90 $\%$ C.L.). The experiment continues to collect data, with the end goal of achieving $\times$10 sensitivity on the previous MEG result.

\subsection{$\mu^{+} \rightarrow e^{+} e^{+} e^{-}$}

In this scenario, the signal is a three co-planar charged tracks, $e^+, e^+$ and $e^-$, coincident in time, and originating from a common vertex and with a total energy equal to the muon mass and with a total momentum of zero. Internal conversions that result in $e^+ e^+ e^-$ and two outgoing neutrinos present a background. The three charged tracks emanate from a common vertex and they are coincident in time. However, the total energy of the tracks is not equal to the muon mass and their momentum is not conserved. Accidental backgrounds also occur, in this case the tracks do not have a common vertex, they are not coincident in time and the kinematics of the charged tracks is different from the signal.

\begin{figure}[ht!]
\centering{
\includegraphics[width=0.45\textwidth]{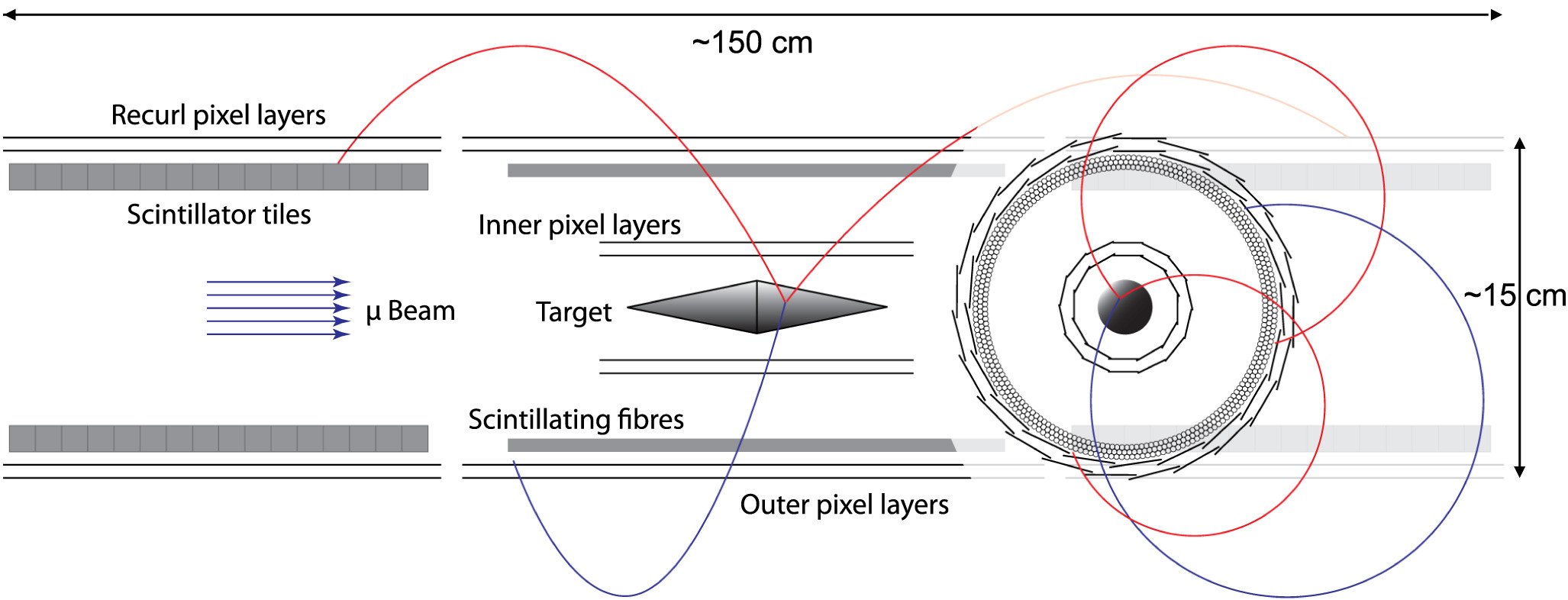}
\caption{A schematic of the Mu3e apparatus. A cone shaped target helps spread out the decay vertices. Tracks a recurled to improve momentum resolution\label{fig:Mu3e}.}}
\end{figure}

Mu3e, also based at PSI, plans to operate in two ``Phases." In Phase-I, Mu3e will use a $\mu^{+}$ beam at a rate of $10^{8}$ $\mu$/s. These muons are stopped on a double-cone target. A schematic of the Mu3e apparatus is shown in Fig.~\ref{fig:Mu3e}. A magnetic field is applied to aid reconstruction of the charged tracks that are then measured using a series of pixel sensors (for position), scintillating fibers and tiles (for timing). At the core of the detector design is a set of 2844 custom monolithic active pixel sensors (HV-MAPS). The pixels are small and thin in order to yield precise position measurements along the electron/positron trajectories and to minimize multiple scattering and energy loss. The pixels consist of 50 $\mu$m of active material and 50 $\mu$m of support structure with an area of 80 $\times$ 80 $\mu$m$^2$.  After leaving the target, the electron/positrons immediately intersect inner pixel layers followed by a set of scintillating fibers which provide an initial time measurement at the pixel layers. The electron/positron tracks are recurled and intersect another pixel layer and scintillating tiles this additional information helps provide the final precise timing measurement.

Phase-I data-taking will begin shortly. Phase-I will yield enough data to provide a factor of $\times$ 100 improvement on the current limit. Phase-II will reach a sensitivity of $2 \times 10^{-16}$ at the 90$\%$ C.L. To achieve this a higher beam rate of $2 \times 10^{9} \mu$/s is required as well as additional pixel recurl layers and scintillator layers to aid in the combinatoric issues at even higher beam rates and to improve the resolutions. 

\subsection{$\mu^{-} N \rightarrow e^{-} N$}

In this scenario, negatively charged muons are captured in the field of the target (Al) nucleus, forming a muonic atom. The muon cascades down in energy to the muonic $1s$ bound state where coherent, neutrinoless muon to electron conversion results in the emission of a mono-energetic electron of energy:
 
 \begin{equation}
     E_{\mu e} = m_{\mu} - E_{BE,1s} - E_{recoil},
 \end{equation}
where $m_{\mu}$ (105.66 MeV/c$^{2}$) is the muon mass, $E_{BE,1s}$ is the binding energy of the $1s$ state, and $E_{recoil}$ denotes the nuclear recoil energy; $E_{\mu e}$ is nuclear dependent, for instance, $E_{\mu e}$ = 104.97 MeV for aluminum (Al).

The Mu2e \cite{mu2eTDR}, at Fermilab, and COMET \cite{COMET:2018auw}, at J-PARC, experiments will search for the coherent, neutrinoless conversion of muon into electron in a muonic atom (\MuToEmConv). Both experiments are based upon a concept proposed in Ref. \cite{MECO}. Experiments searching for \MuToEmConv~ have the advantage of being free of accidental backgrounds, however, there are two general classes of backgrounds that must be eliminated:
\begin{itemize}
    \item Backgrounds that scale with the number of muons. 
    \begin{itemize}
        \item \textbf{Decay-in-orbit (DIO)}: arises when muons in the 1$s$ state undergo normal weak decay while in orbit: $\mu^- \rightarrow e^- \nu_{\mu} \bar{\nu}_e$. The energy, $E_{\mu e}$, of the outgoing conversion electron is well above the end-point energy of the free muon decay spectrum ($\sim 52.8$ MeV) but when the decay occurs in the presence of a nucleus the outgoing electron can exchange momentum with the nucleus, resulting in a long recoil tail \cite{czarnecki}. Experiments searching for \MuToEmConv~ must design their tracking detectors for high momentum resolution to mitigate against these tail electrons. For example, Mu2e employs a straw tube tracker made of 20,000, 5 mm diameter thin straws orientated in sets of panels. These panels are oriented in circles ($3 \times 120$ degrees) and along the beam axis, and combined in sets to form panels, two panels form a station. The overlap of straws at varying angles allows for stereo reconstruction. This geometry, along with sophisticated reconstruction algorithms, allows for momentum resolution $<$ 180 MeV/c. In addition, the center of the tracker is kept hollow, already completely blinding the detector to low momentum ( $\lesssim 75 $ MeV/c) electrons from the target.
        \item \textbf{Radiative Pion Capture (RPC)}: arises from the process, $\pi^-N \rightarrow \gamma N^{\prime}$, where a $\sim$ 105 MeV/c electron can be produced from the photon pair-production.  Both Mu2e and COMET use a pulsed proton beam to overcome this background.  Pions have a short (26 ns) lifetime compared to muons (2.2 $\mu$s) and the experiments employ a delayed live-gate to effectively ``wait out" pion decays.  Mu2e will utilize a time-window of $\sim$ 700 ns in the 1695 ns period of the Fermilab Delivery Ring and COMET will use a time window from $\sim$ 700 to 1170 ns.
    \end{itemize}
    \item Backgrounds that scale with running time.
    \begin{itemize}
        \item \textbf{Cosmic-ray Induced} : cosmic muons can knock electrons out of the stopping target or provide other sources of backgrounds which produce $\sim$ 105 MeV/c particles. Neutral secondaries from cosmic rays can also generate backgrounds. Both Mu2e and COMET employ active cosmic ray veto systems to eliminate cosmic-ray induced backgrounds.
    \end{itemize}
\end{itemize}

\begin{figure}[ht!]
\centering{
\includegraphics[width=0.45\textwidth]{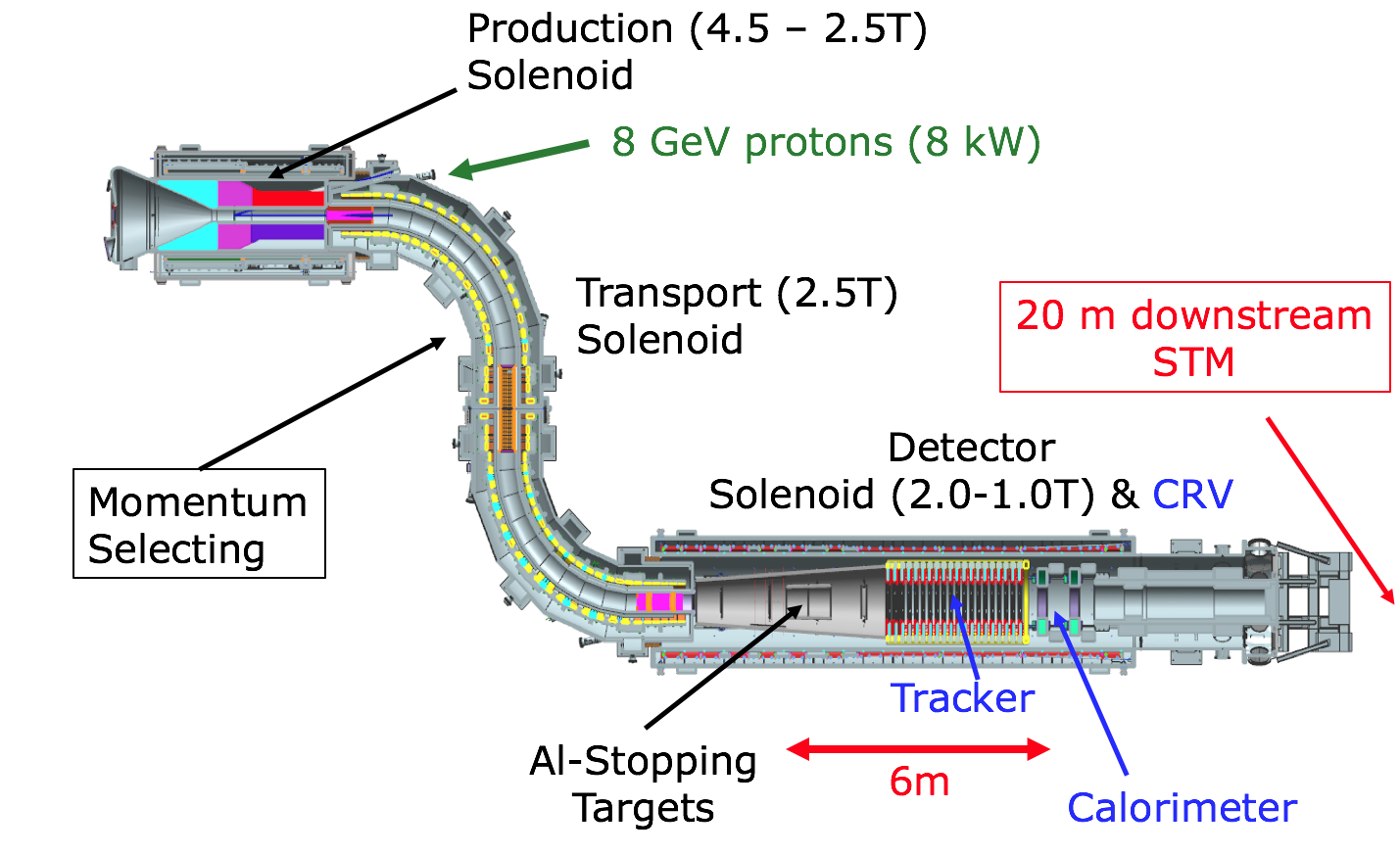}
\caption{A schematic of the Mu2e apparatus \label{fig:mu2e}. An intense muon beam is created in the first of three solenoids. The $\mu^{-}$ with low momentum are transported to the stopping target using a graded field.}}
\end{figure}

\begin{figure}[ht!]
\centering{
\includegraphics[width=0.3\textwidth]{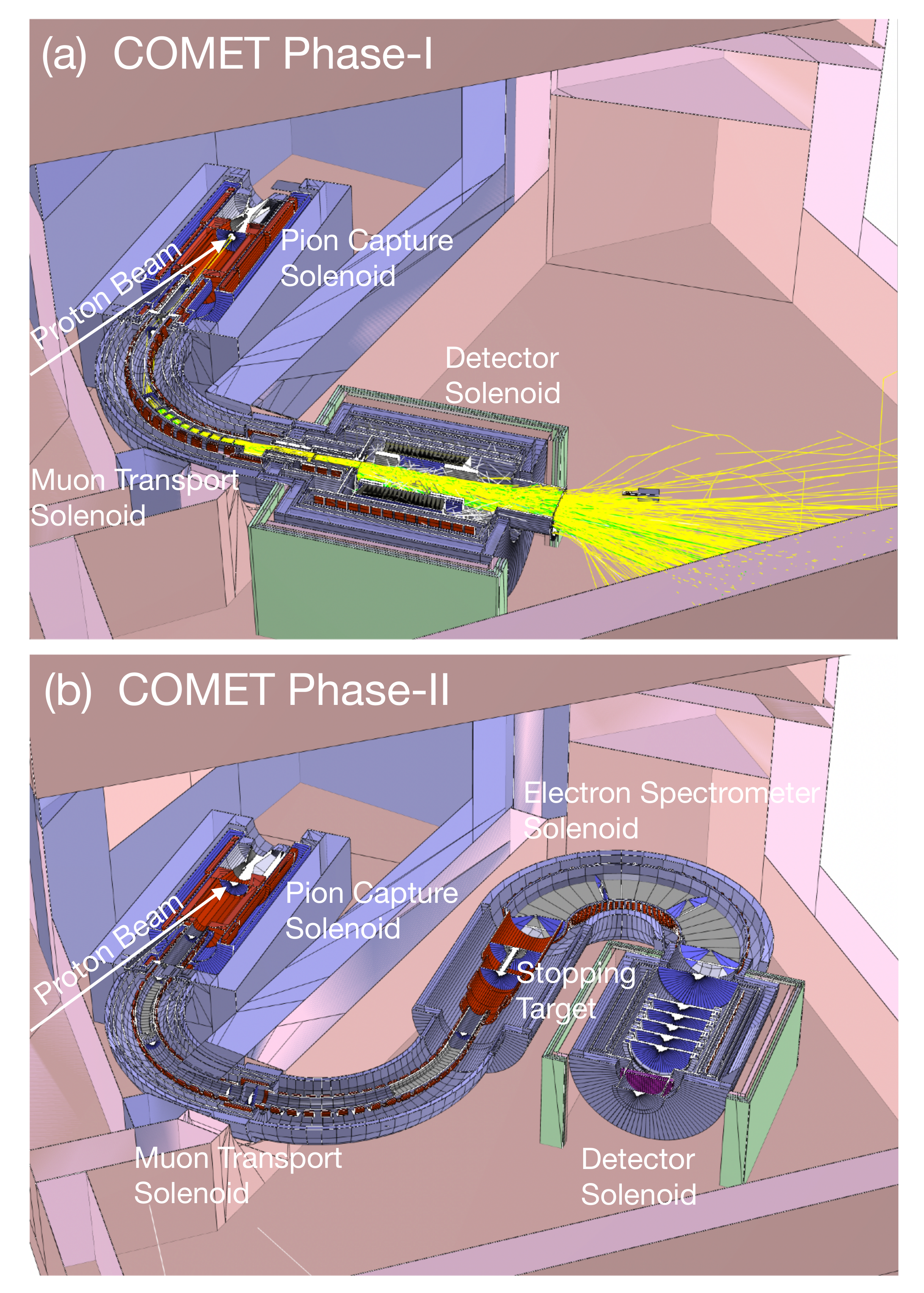}
\caption{A schematic of the COMET apparatus. Phase-I and II are distinctly different.  There are several difference from Mu2e, most notably the placement of the stopping target and the fact the a spectrometer is used in a final bend before the detectors \label{fig:COMET}.}}
\end{figure}

The Mu2e and COMET apparatus are shown in Fig.~\ref{fig:mu2e} and ~\ref{fig:COMET} respectively. Mu2e and COMET are designed to overcome the limitations faced by SINDRUM-II. In addition to eliminating all backgrounds by design both will employ an intense muon beam that is created and transported via a series of super-conducting solenoids with graded magnetic fields. 

The Mu2e experiment muon beam is produced as a tertiary beam. 8 GeV, 8 kW protons intersect a tungsten target inside the production solenoid. The experiment collects the backward moving (with respect to the incoming beam) $\pi^{-}$ that decay into $\mu^{-}$. The transport solenoid has an ``S" shape that helps remove line of site backgrounds. The transport solenoid contains collimators that select negative charged, low momentum muons. Antiproton windows remove any remaining $\bar{p}$'s in the beam. The detector solenoid is the final solenoid and contains the aluminum stopping target, consisting of 37 thin ($\sim$100$\mu$m) aluminum foils (to minimize scattering and energy losses from the outgoing conversion electron). The magnetic field in the detector solenoid results in helical electron tracks. The momentum of these tracks is measured by the aforementioned straw tube tracker followed by a CsI calorimeter system which helps identify the electrons.

COMET will take place across two distinct ``Phases." The end geometry is similar to Mu2e's with three distinct super-conducting solenoids. Phase-I utilizes a 3.2 kW, 8 GeV proton beam which produces muons as tertiary decay products. In Phase-I, the capture solenoid, the transport solenoid, and the detector solenoid are present, however, the transport solenoid is constructed only up to the end of the first 90 degree bend. Phase-I will allow for muon beam measurements as well as an intermediate conversion search with sensitivity of $\mathcal{O}(10^{-15})$. The muon beam measurement will be carried out with the StrECAL detector system, composed of straw-tube trackers and an electron calorimeter. The conversion search will be performed with the CyDet system, which consists of a cylindrical drift chamber (CDC) and cylindrical trigger hodoscopes (CTH) surrounding the stopping target. The CyDet system replaces the StrECAL once the muon beam measurement is carried out. Muons stop in the aluminum target discs placed at the center of the detector solenoid. The transversely-emitted electrons from the muon stopping target are detected with the CDC and hit the CTH. In Phase-II, the beam power is 56 kW. The transport solenoid is completed. The electron spectrometer solenoid (ES)  is a 180 degree bend curved solenoid installed at the downstream of the muon stopping section of the transport solenoid. It selects the charge and momentum of the emitted electrons with its curved solenoidal structure. Low momentum electrons are eliminated by the DIO blocker which is installed at the end of the ES. The StrECAL system will be upgraded and used as the main detector in Phase-II.

Mu2e will begin collecting data in early 2027. The full data sample will be taken over two run periods. Ref.~\cite{SU2020} details the expected upper limit at 90 $\%$ C.L and $5 \sigma$ discovery potential of Run-I to be $6 \times 10^{-16}$ and $1 \times 10^{-15}$ respectively. The total sensitivity of Runs I and II is projected to be $\mathcal{O}(10^{-17})$. The hardware employed in both runs will be identical.  COMET Phase-I is intended to begin in 2025 and has a projected upper limit of $5 \times 10^{-15}$ at 90 $\%$ C.L. Phase-II is a significant upgrade from Phase-I and is expected to be complete a few years after the completion of Phase-I. The sensitivity of Phase-II is projected to be comparable to that of the full Mu2e program.

\subsection{Looking Further Ahead}

The next decade will see unprecedented improvements in our understanding of muon CLFV. In the event of a signal in any of the outlined CLFV searches, this would open a new age of particle physics and signify, conclusively that beyond Standard Model physics was at play. Attention would then turn to understanding the observed signal through subsequent measurements. Mu2e-II \cite{Mu2eII} is a proposed extension to Mu2e which could be built at Fermilab utilizing the PIP-II beam. Much of the Mu2e hardware would be reused. The projected sensitivity of Mu2e-II is $3 \times 10^{-18}$. R$\&$D continues in order to finalize the detector designs for Mu2e-II. In addition, there is an ambitious proposal to build an Advanced Muon Facility (AMF) \cite{AMF} at Fermilab. This machine would host a suite of CLFV searches, searches for all three muon CLFV channels could be conducted. The facility would also provide intense muon beams for an array of additional studies. The design of AMF is in its infancy and a large amount of R$\&$D is required to complete its conceptual design.

\section{Conclusion}

The current decade has already seen high-profile results from the Fermilab g --2 Experiment which provided a measurement of $a_{\mu}$ to world-leading precision. The effort to resolve tensions in the SM predictions continues, with the J-PARC  g --2 Experiment expected to provide a complementary measurement of $a_{\mu}$ in the coming years. Likewise, searches for muon CLFV are a powerful probe for new physics, with already tight constraints, orders of magnitude below analogous constraints on tau processes. In the coming decade results from MEG-II, Mu3e, COMET and Mu2e will collectively explore an array of new physics operators and are expected to constrain new physics models up to $10^{4}$ TeV/c$^{2}$.

\bibliographystyle{utcap_mod}
\bibliography{paper}

\end{document}